\documentstyle[12pt,preprint]{aastex}	% for preprint

\def\gradip{\hbox{\rlap{\hbox{.}}\raise 5.truept \hbox{{\small $\circ$}}}}

\catcode`\@=11
\def\gsim{\ifmmode{\mathrel{\mathpalette\@versim>}}
    \else{$\mathrel{\mathpalette\@versim>}$}\fi}
\def\lsim{\ifmmode{\mathrel{\mathpalette\@versim<}}
    \else{$\mathrel{\mathpalette\@versim<}$}\fi}
\def\@versim#1#2{\lower 2.9truept \vbox{\baselineskip 0pt \lineskip
    0.5truept \ialign{$\m@th#1\hfil##\hfil$\crcr#2\crcr\sim\crcr}}}
\catcode`\@=12

\begin{document}

\title{Testing intermediate-age stellar evolution models with VLT photometry
of LMC clusters. III. Padova results\altaffilmark{1}}

\author{
Gianpaolo Bertelli\altaffilmark{2,3},
Emma Nasi\altaffilmark{3},
Leo Girardi\altaffilmark{4},
Cesare Chiosi\altaffilmark{5},
Manuela Zoccali\altaffilmark{6},
Carme Gallart\altaffilmark{7,8}
}

\altaffiltext{1}{Based on observations collected at the European Southern
Observatory, Paranal, Chile.}
\altaffiltext{2}{National Council of Research, IASF-CNR, Rome, Italy; bertelli@pd.astro.it}
\altaffiltext{3}{Osservatorio Astronomico di Padova, Vicolo dell'Osservatorio 5, 35122 Padova, Italy; nasi@pd.astro.it}
\altaffiltext{4}{Osservatorio Astronomico di Trieste, Via Tiepolo 11, I-34131 Trieste, Italy; lgirardi@ts.astro.it}
\altaffiltext{5}{Dipartimento di Astronomia dell'Universit\`a di Padova, Vicolo dell'Osservatorio 5, I-35122- Padova, Italy; chiosi@pd.astro.it}
\altaffiltext{6}{European Southern Observatory, Karl Schwarzschild Strasse 2, D-85748 Garching bei M\"unchen, Germany; mzoccali@eso.org}
\altaffiltext{7}{Andes Prize Fellow,  Universidad de Chile and Yale University}
\altaffiltext{8}{Ramon y Cayal Fellow. Instituto de Astrofisica de Canarias, 38200 Tenerife, Canary Islands, Spain; carme@iac.es }
%________________________________________________________________________

\begin{abstract}

The  color-magnitude diagrams (CMDs) of three intermediate-age LMC clusters, 
NGC 2173, SL556 and NGC2155 are analyzed to determine their age and 
metallicity basing on Padova stellar models. 
Synthetic CMDs are compared with cluster data. The best match 
is obtained using two fitting functions based on star counts in the different
bins of the cluster CMD. Two different criteria are used.
One of them takes into account the uncertainties in the color of
the red clump stars. Given the uncertainties on the experimental values
of the clusters metallicity, we provide a set of acceptable solutions.
They define the correspondent values of metallicity, age, reddening and 
distance modulus (for the assumed IMF). The comparison with Padova models
suggests for NGC 2173 a prolonged star formation (spanning a period of 
about 0.3 Gyr), beginning  1.7 Gyr and ending  1.4 Gyr ago. The metallicity Z
is in the range 0.0016 $-$ 0.003.
Contrary to what suggested for NGC 2173, a period
of extended star formation was not required to fit the SL 556 and the NGC 2155
observations.  
For SL 556 an age of 2.0 Gyr
is obtained. The metallicity value is in the range 0.002 $-$ 0.004, depending
on the adopted comparison criterium.
The derived age for NGC 2155 is 2.8 Gyr and its metallicity Z is in the 
range 0.002 $-$ 0.003. The CMD  features of this 
cluster suggest that a more efficient overshoot should be adopted 
in the evolutionary models.

\end{abstract}

%________________________________________________________________________

\section{INTRODUCTION}
\label{intro}

The Magellanic Clouds are a natural target to study the richest 
resolvable clusters in the Local Group because of their proximity.
They provide information on  stellar and chemical evolution, on stellar
populations
and on their star formation history (Da Costa 2001, 
Rich et al. 2001, Keller et al. 2001, Holtzman et al. 1999, Bica et al.
1998, Santos et al. 1999, Dirsch et al. 2000, Hill et al. 2000).

Despite the great achievements of the stellar evolution theory there are some 
points of disagreement between theory and observations. They are 
related to our poor knowledge of the extension of convectively unstable regions
and associated mixing processes. A number of observational facts 
suggest that in stellar interiors the convective regions extend beyond the
classical Schwarzschild limit (for example, the width of the main sequence 
band, the 
number ratio of stars in different evolutionary phases in Hipparcos data
or in  intermediate age and old open star clusters, and other data as 
discussed by Chiosi 1999).
Several effects can produce an extension of the convective core above that
predicted by classical models. 
Recent studies indicate that rotation can provide a natural 
way to increase internal mixing and hence a larger core
(Heger et al. 2000, Meynet \& Maeder, 2000).
Some recent papers are suggesting that the CMD morphology of populous 
young
and intermediate-age clusters can only be matched by stellar models  if a
significant amount of convective core overshoot (CCO) is assumed (Rosvick
\& VandenBerg 1998, Keller, Da Costa \& Bessell 2001).  
The use of OPAL opacities (Iglesias \&
Rogers, 1993) resulted in a substantial increase in the size of the
classical core.

The comparison of observed cluster CMDs with the predictions of stellar 
evolutionary models is the usual way of testing the different parameters
and assumptions used to compute them (Chiosi et al. 1989, Lattanzio et al.
1991, Vallenari et al. 1992, Barmina et al. 2002).
 In this paper we will use current 
Padova stellar evolutionary models (Girardi et al. 2000) to analyze the data 
of three LMC clusters described in {\bf Paper I} (Gallart et al. 2002). 
In the context of a common project aiming to test stellar evolution comparing
models to accurate LMC clusters data, 
the data and a preliminary analysis are presented in the first paper of this 
series. 
Paper II, based on Yonsei-Yale stellar models, and Paper III on  Padova models,
present  completely independent and detailed analysis of the same clusters
with the main goal of testing the need and the efficiency of convective 
overshoot in the LMC range of age and metallicity, in addition to the
differences of the two evolutionary codes.

In this paper we have 
developed a  comparison method using synthetic CMDs,
first evaluating the distance modulus and reddening, and subsequently 
determining the best age and metallicity combinations. 
We further assess whether current stellar evolution models, in particular 
the amount of convective core overshoot  assumed in them, are
adequate to reproduce the characteristics of the analyzed sample of
intermediate-age low-metallicity Large Magellanic Cloud clusters. Note
that thus far most comparisons were performed on Galactic open 
and globular clusters, which are, respectively, young and intermediate-age
solar metallicity, or old metal-poor populations.
The intermediate-age LMC clusters have in general metallicities which are
lower  than those of Milky Way clusters of the same age. They provide a way 
to test a new area of the parameter space.   

In section 2 the input physics of Padova models and isochrones is 
described, together with the efficiency and constraints adopted for the
extension of the convective core, pointing out the more significant
differences with respect to the Yonsei-Yale models ($Y^2$ by Yi et al. 2001).  
In section 3 the method of synthetic diagrams is discussed. The comparison
method between simulations and data is presented in section 4 and the results
for the three clusters are described in sections 5, 6 and 7. Section 8
presents the final discussion and the summary of the results.

\section{EVOLUTIONARY MODELS AND ISOCHRONES}

The considered evolutionary models are from Bertelli et al. (1994) and
Girardi et al. (2000). In the following the input physics of Padova
models and the adopted assumptions for overshoot are summarized in order to
give a clear overview of the more significant differences with respect to 
$Y^2$ evolutionary models (Yi et al. 2001).   

The Padova stellar models are evolved
from the ZAMS to the whole H- and He-burning phases  for initial masses 
between $0.15-7 M_{\odot}$. In the isochrones
the TP-AGB phase for intermediate and low mass stars is included as in
Girardi et al. (2000).
 More massive models by Bertelli et al. (1994)    
are used to complement those by Girardi et al. (2000) for younger
isochrones as their input physics is coherent with that of new lower
mass models.

The stellar models are evolved to the onset of helium burning in the 
$Y^2$ case for the mass range $0.4-5.2 M_{\odot}$ and 
the ages  are computed starting from the deuterium
MS, also called  the "stellar birthline" during the pre-MS stage.
The pre-MS lifetime (from the birthline to the ZAMS) depends on metallicity
and is a strong function of 
stellar mass, varying from less than 1 Myr for a 5 $M_{\odot}$ to about 
200 Myr for a 0.4 $M_{\odot}$ (Yi et al. 2001). For a solar mass model it is 
about 43 Myr long. Because of this, isochrones considering ZAMS stars as 
"zero-age" are bound to mismatch the lower part of the observed CMDs of
young clusters that contain pre-MS stars. This effect is negligible in
old ($> 1$ Gyr) populations, causing age underestimates of less than $ 1 \%$ for
populations from 0.1 to 10 Gyr for the considered metallicities (Yi 2002,
private communication).   

Differences in the input physics of the two sets of models (Padova and 
$Y^2$) are: 

$\bullet$ the equation of state: MHD EOS (Mihalas et al. 1990) and Straniero (1988)
for Padova models,  OPAL EOS (Rogers et al. 1996) for $Y^2$ models;

$\bullet$ the energy generation rates: Caughlan \& Fowler (1988) and Landr\`e et al.
(1990) for Padova models, Bahcall \& Pinsonneault (1992, 1994 private 
communication) for $Y^2$ ones;   

$\bullet$ the  mixing length parameter: $l/H_p = 1.68 $ (Padova) and = 1.743 
($Y^2$);

$\bullet$ Helium diffusion: according to Thoul et al. (1994) for Yi et al. (2000),
 not included in Girardi et al. (2000), even if diffusive mixing was taken 
into account by Salasnich (2000);
    
$\bullet$ galactic helium enrichment parameter: the ratio $\Delta Y / 
\Delta Z $ was adopted equal to 2.25 for Padova models and to 2.0 for $Y^2$ 
ones.

Most of the quoted points do not produce significant differences in the 
evolutionary behaviour of stellar tracks in the theoretical HR diagram 
(see Figure 6 in Paper I). 
A different helium abundance for given metallicity is the most important
cause of differences between models of various authors.

The treatment of convective core 
overshoot and the conversion from the theoretical to the observational
plane deserve a more detailed description.

\subsection{Convective Overshoot}

The argument for the occurrence of convective overshoot is that the traditional
criteria for convective stability look for the locus where the buoyancy
acceleration vanishes. Since it is very plausible that the velocity of the 
convective elements does not vanish at that layer, they will penetrate
(overshoot) into regions that are formally stable. Although the physical 
origin of
convective overshoot is simple, its efficiency and description are much 
more uncertain. This uncertainty is reflected in the variety of solutions
and evolutionary models that have been proposed over the years. 
Major contributions to this subject in the past were from Shaviv \& Salpeter 
(1973), Prather \& Demarque (1974),
Maeder (1975), Cloutman \& Whitaker (1980), Bressan et al. (1981), Stothers
\& Chin (1981). See Chiosi (1998) for an exhaustive review for the 
understanding of the HR diagrams and for references on convection, overshoot 
and related stellar models.

The physics of convection has made significant progress, but due to its
complexity it is not suitable for the computation of large sets of evolutionary
tracks.  The vast majority of stellar models are computed with the simple
parametric description of the overshoot extent in terms of the pressure scale
height (for example Geneva libraries, Schaerer et al. 1993, Charbonnel et al.
1999 and references 
therein), or adopting a ballistic description of convective motions (Bressan
et al. 1981, Girardi et al. 2000, Padova library).
As far as convective overshoot 
is concerned the problem can be split in two parts: 1) how far the convective 
elements can penetrate into the radiative regions above the border set by the
Schwarzschild criterion; 2) how efficiently this material is mixed with the
surrounding matter (Deng et al. 1996).

In particular the extension of convective regions in Padova models is 
estimated according to the formalism by Bressan et al. (1981). It makes
use of the mixing length theory (MLT), and therefore contains the parameter 
$\Lambda_c = l / H_p$, where $l$ is the mean free path of convective elements
and $H_p$ is the pressure scale height.  
We remind that according to this formalism the parameter $\Lambda_c$
 expresses  the overshoot distance 
{\bf across} the Schwarzschild border  in units of the pressure 
scale height. Other authors define that distance as
measured {\bf above} the convective border (for example the Geneva group).
This means that $\Lambda_c = 0.5$ in Padova models roughly corresponds
to a 0.25 pressure scale height for  Geneva and $Y^2$ models.

In Girardi et al. (2000) the following prescription was adopted for the 
parameter  
$\Lambda_c$  as a function of stellar mass  during the H-burning phase:

$\bullet$ $\Lambda_c$ is set to zero for stellar masses $M \leq 1.0 
M_\odot$ (the core is radiative).

$\bullet$ In the range between 1. and 1.5 $M_\odot$ the overshoot efficiency
gradually increases with mass ($\Lambda_c = M/M_\odot - 1.0)$.
In this mass range the overshoot efficiency is still uncertain and this is 
the mass range that we will investigate in this paper.

$\bullet$ For $M \geq 1.5 M_\odot$ the value $\Lambda_c = 0.5$ is adopted.

In the core helium burning stage the value $\Lambda_c = 0.5$ is used for all
stellar masses.

Overshoot at the lower boundary of convective envelopes is also considered
as described in Girardi et al. (2000). It is smaller for lower masses and more 
significant  for masses greater than $M > 2.5 M_\odot$ as in Bertelli et 
al. (1994). In the mass range involved in the following discussion (lower
than 1.5 $M_\odot$) envelope overshoot does not significantly change  the
evolution in the HR diagram.

Woo \& Demarque (2001) discuss the empirical constraints on convective overshoot
and, following Roxburgh's integral constraint, they implement an upper limit of
overshoot within the conventional method of parameterization to remove an overly
large overshoot effect on low mass stars.
Yi et al. (2001) adopted a moderate overshoot (OS=0.2 $H_p$) in their 
evolutionary computations for younger
isochrones ($\leq 2$ Gyr) and OS=0.0 for older ones ($\geq 3$ Gyr).

\subsection{Conversion from ($Z, L, T_{eff}$) to CM diagrams}

Most of the differences  between Padova and $Y^2$ isochrones (see section 5 of 
Paper I, Gallart et al. 2002) can be ascribed to the transformation  
of the  theoretical quantities (Z, L, $T_{eff}$ into colors and magnitudes,
due  to  the large uncertainties in the model 
atmospheres for  cool stars and giants.  

Yi et al. (2001) used the color transformation tables of Lejeune, Cuisinier 
\& Buser
(1998) which are based on the Kurucz (1992) spectral library, substantially 
modified to better match the empirical stellar data of solar metallicity 
and extended to low 
temperatures based on empirical data and low temperature stellar models.

Also Bertelli et al. (1994) and Girardi et al. (2000) isochrones are based 
on the Kurucz library of stellar spectra and are extended at high temperature 
with black-body spectra. At low temperatures the spectral distribution of late
type stars was derived from three different sources (Lancon \& 
Rocca-Volmerange, 1992, Terndrup et al. 1990, 1991 and Straizys \& 
Sviderskiene, 1972). They were combined in such a way to reproduce the 
observed value of the colors (V-K) and (J-K) for the considered  spectral type. 
A detailed description of the assemblage of the conversion relationships
can be found in Bertelli et al. (1994).

The effects of the different color transformations on the [V,(V-R)] CMD
are more pronounced than the effects of the different input physics on the
location of the isochrones in the theoretical plane. This can be seen in 
Figure 6 of Paper I, where in the observational plane it can be noticed that
the Padova isochrones are slightly redder (by $\simeq 0.03 mag$) than the
$Y^2$ ones along the main sequence and on the RGB, and more significant
differences are present  in the turn-off region, the last effect being a
combination of both the different input physics and the color transformations.  
From the comparison of our results with those by Woo et al. (2002, Paper II)
it appears  that (for the same initial conditions: distance modulus and 
metallicity) ages based on Padova models are 0.1 Gyr younger than those 
obtained by Woo et al. (2002).
 
\section{SYNTHETIC CMDs}

The method of synthetic diagrams is used to compare the distribution of
stars in different regions of the observed and the simulated CMD.
This method allows 
for the  evaluation of different hypotheses in order to determine which one 
better reproduces the observed features.

Starting from evolutionary models, we distribute stars in the CMD
defining age and metallicity (or a range of ages and metallicities),
adopting a value for the reddening, the distance modulus to the LMC cluster 
and an initial mass function (IMF) slope.

The synthetic diagrams are constructed by means of a MonteCarlo algorithm, which
randomly distributes stars in the CMD  according to 
evolutionary lifetimes and given the initial mass function slope. 
The code takes into account completeness effects and the photometric errors 
affecting the observational data, by adding
an artificial dispersion to the magnitudes and colors of the simulated stars.
The generated CMDs can take into account the effects of an age spread at 
the time of formation of the cluster population, the different slopes of the IMF, 
and the presence of a certain fraction of binary stars. 
We populate our synthetic CMD until an assigned
number of main sequence stars $N_{MS}$ is matched, corresponding to
the total number of MS stars present in the data. 

\subsection{Adopted parameters}

There are a few parameters to be fixed in the simulations:
the metallicity, the age (and if necessary an age dispersion), the reddening, 
the distance modulus and the initial mass function (IMF).

\noindent
$\bullet$ The IMF

The IMF is defined by
\begin{equation}
 dN \quad \propto \quad M^{-\alpha}dM
\end{equation}

\noindent
for which we initially assume the classical Salpeter law
with $\alpha =2.35$.  We also consider another power law value ($\alpha =1.35$).
The normalization parameter is fixed by the number of main sequence stars
in the observed CMD.

\noindent
$\bullet$ Chemical composition.
 
There are determinations of the metallicity of the three clusters from
photometry, from Ca II triplet spectroscopy, or other methods, and 
these will be quoted in the description of each cluster analysis.
As a starting point for this study we explored values of Z in the range
$0.001 \le Z \le 0.004$, which includes the empirical determinations.
The adopted metallicities derive from the synthetic CMD best matching the
observations for various ages.

\noindent
$\bullet$ Distance modulus

The distance modulus (DM) to the LMC is the subject of a long-lasting 
controversy
with a difference of more than 0.5 mag between the extreme values.
The HST Key Project on the extragalactic distance scale (Freedman et al. 2001)
has adopted DM= 18.5 $\pm$ 0.1 mag based on revised Cepheid distances
and Girardi \& Salaris (2001), taking into account population effects on the 
red giant clump absolute magnitude, obtained 18.55 $\pm 0.05$.
Benedict et al. (2002) provide a weighted average for the distance modulus 
to the LMC (DM = 18.47 $\pm$ 0.04) based on absolute 
parallaxes and relative proper motions of RR Lyrae (from FGS 3 on HST) 
and an exhaustive list of recent LMC distance modulus determinations.       

Given that each of the observed clusters can be located at distances slightly 
different than the bulk of the LMC stars, in our simulations we allowed the
DM to vary and for each cluster its value is obtained from the comparison
between data and synthetic CMDs.  

\noindent
$\bullet$ Reddening

The reddening is known to be patchy and to vary from one region to the other of
the LMC. A large number of different estimates exist in the literature
with values in the range from 0.03 to 0.22 mag, depending on the region and 
on the reddening indicator adopted. Since reddening is a major ingredient in
any distance determination procedure, it is important to take into account the
close interdependence of the variations in distance and those in reddening,
when trying to simulate the observations.

Larsen et al. (2000) estimate the reddening in the LMC from Stromgren CCD
photometry, finding a foreground reddening of E(B-V)=$0.085 \pm 0.02$,
while for Schlegel et al. (1998)  typical reddenings of E(B-V)=0.075
toward the LMC are estimated from the median dust emission in surrounding 
annuli.    
The reddening distribution  has also been derived  by 
Oestreicher et al. (1995). They found a mean foreground reddening of 0.05  
$\pm 0.02$ mag.   

In our simulations we considered a reasonable range of reddening values
and adopted that better reproducing  each cluster data.

\noindent
$\bullet$ Binaries

The overall frequency of binaries, defined as the fraction of primaries that
have at least one companion, is at least 50 percent (Duquennoy \& Mayor 1991).
The binary fraction appears to increase ($\sim 70$ percent) with increasing  
primary mass, at least among the more massive stars, and to decrease to around
30-40 percent for M stars (Mayor et al. 2001, Larson 2001).
When considering binaries in the CMD simulations, it is necessary to assume a 
distribution of mass ratios $ q = M_2 / M_1$ between the secondary and the 
primary component of the system. Mermilliod et al. (1992) found a 
mass-ratio distribution  reasonably flat in the range $ 0.4 < q <1$ and no
information for $q < 0.4$ from spectroscopic binaries in the Pleiades cluster.
The main effect of unresolved binary stars on the CMD of stellar clusters 
is the appearance of a second sequence, running parallel to the main sequence
with brighter luminosity (up to $\sim$ 0.7 mag for binary components of equal 
mass) and cooler temperature.
Such a sequence can be observed in the CMDs of NGC2173 and SL 556.

In our simulations we adopt the binary fraction as described for each cluster,
and the mass ratio distribution from Mermilliod et al. (1992), neglecting the
low mass-ratio companions, since the magnitude and color of these binaries 
(with the primary rather more massive than the secondary) are
almost the same as those of a corresponding single star.

\section{ COMPARISON METHOD}

We divide  the CMD in a grid with bins of 0.05 mag in color and of
0.2 mag in V magnitude, and  compare the distribution of stars in the synthetic 
and in the observed CMD by minimizing  suitable functions.
We consider two fitting functions, linked to the number of stars in each bin 
of the grid, defined in the following way:

\begin{equation}
f1 = 1/N_{o} \sum_{i=1}^n \big[ (N_{mod} - N_{obs})^2 / (N_{mod} + N_{obs})\big]_i 
\end{equation}
\begin{equation}
f2 = 1/N_{o} \sum_{i=1}^n \big[ (N_{mod} - N_{obs})^2 \sqrt(N_{mod} + 
N_{obs})\big]_i   
\end{equation}

$N_{mod}$ and $N_{obs}$ are the number of objects in each bin of the 
simulated and the observed diagram respectively.
$N_o$ is the number of bins of the observed diagram that are not empty and 
the sum is extended to all bins where there are observed and/or simulated
stars. The more the distributions are similar, the lower the function value.
Whereas $f1$ gives the same weight to every bin, $f2$ gives more weight to more
populated bins, so a comprehensive view of the results takes into account
information from both functions.
The functions as defined in this paper are preferred over a (weighted)
$\chi^2$ function. The main reason is that a $\chi^2$-fitting function
does not sufficiently take into account the non-matching simulated points.
These points contain important information for a proper overlay of the 
observed with the synthetic CMD (see Ng et al. 2002).

 The analysis looks for the minimum of the two functions $f1$ and $f2$
varying the input parameters of the simulated CMD, i.e.
metallicity and age, for given IMF slope and binaries percentage with the value 
of the distance modulus and the reddening determined previously with a series
of synthetic CMDs. A better agreement between the simulation and the observed 
data is obtained with a lower value of the fitting functions.
They get a maximum value for two non-overlapping CMDs,
but are at  minimum when a better overlay is obtained.
In order to minimize the effects of  statistical fluctuations of stochastic 
nature  several synthetic 
diagrams were generated with the same input parameters for each considered 
case in the Monte Carlo simulations. The minima from the average values of 
the fitting functions are derived 
and considered as the best solution, since the minimum derived from
the average  values of many simulations is more reliable than 
a single minimum (see section 5.1 and footnote 10). 

There are uncertainties in the observed color, related to the 
photometric calibration and the estimated reddening, in addition to those 
in the V magnitude, linked to the magnitude calibration and  the 
distance modulus determination. To take them into account for each simulation we
allow  the zero point of the CMD to vary inside a box  whose sides ($\Delta c, 
\Delta v$) are twice the evaluated uncertainties in color and in V magnitude. 
In such a way  it is 
possible to evaluate the peculiar values of the displacements of the zero 
point that gives the best agreement between the simulated and the observed 
diagram. This is again obtained via the minimization of the two functions 
$f1$ and $f2$. 
  
The analysis of the CMD is made by separately
evaluating  the blue region (main sequence stars with $(V-R) 
\le 0.35$ with the related $f1ms$ and $f2ms$ ) and the red region 
(stars with $(V-R) > 0.35$ with the related $f1red$ ). 
The function $f2$ proved to be significant and useful only for blue stars, and 
not for red stars because of the low number of stars in the red region.

\subsection{Adopted criteria} 

The conversion relationships from models to CMD plane are a critical 
ingredient in 
the comparison between theory and observations. They can affect
in a differential way both  the turnoff and  the red clump. 
For example comparing results by Bertelli et al. (1994) with 
those by Girardi et 
al. (2002), who updated the color conversions, the differences are of the order 
of 0.02 magnitudes in the horizontal branch (HB) V-R  color, 
while the luminosity change is negligible,
 and the effect on the main sequence is insignificant.  
Also the slope of the red giant branch is affected more and more 
at decreasing effective temperature of stellar
models (differences between Padova and $Y^2$ CMDs are shown in Paper I). 

For this reason we have not considered red giant branch stars more luminous
than the clump.  The criteria for
the comparison are: i)  the
synthetic models must properly reproduce the main sequence (turn-off
and termination point) and at the same time the color and the luminosity
of the red clump;  ii) the constraint of reproducing the color of the clump 
is relaxed (but
not its luminosity). The first is called "global criterium" since it requires
the fit of the global properties of the cluster and  the second is named 
"partial criterium".

In the following the analysis  will proceed in two steps:

\noindent
{\bf first step}: a series of synthetic CM diagrams are computed for various 
metallicities and inside  an age range suggested by a preliminary analysis 
with the isochrones. The overlap of the simulated to the observed clusters
according to the prescriptions of the global or partial criterium 
defines distance modulus and reddening values;

\noindent
{\bf second step}: a detailed and systematic  comparison of data with simulated 
clusters is done with input parameters (metallicity and age) 
varied  inside a prefixed grid for given binaries percentage and IMF slope. 
It is performed  with the aid of the fitting functions and is      
devoted to a more precise determination of age and chemical composition.

\section{NGC 2173}

Previous results on the metallicity of NGC 2173 report    
values in the range from [Fe/H]=-0.24 to -1.4 and for the age  between 
1.4 and 2.0 Gyr (Cohen 1982,
Bica et al. 1986, Olszewski et al. 1991, Mould et al. 1986, Geisler et al. 
1997). 

Looking at the color magnitude diagram of NGC 2173 (Figure 1, left panel)
we note  the following:

1) stars at the main sequence termination point are scattered in an unusual 
manner. This feature might be produced by various causes, among which in the
following we
will discuss the presence of binaries, the inadequate subtraction of the
field, a differential reddening across the disc and a prolonged star formation. 

2) the number of observed subgiants and RGB stars less luminous than the 
clump looks significantly  lower than that of the RGB stars
more luminous than the clump. This is contrary to what is expected  
from the models.

3) the red border of the MS (at luminosity lower than the TO) is not as wide
as that obtained taking into account a reasonable fraction of binary stars.
However the presence of binaries is suggested by the vertical
structure of the MS termination point.

We tested the effect of different percentages of binary stars on the 
morphology of the turnoff, but there were not significant differences
varying the percentage from $ 30 \% $ to $50 \%$, and it was not sufficient  
to produce a good agreement with the
observed distribution and the shape of the termination point. 
If the statistical subtraction of the field is not adequately performed, it
may introduce a bias in some regions of the observed CMD. We will discuss in
section 5.4 an experiment showing that the field subtraction cannot have
artificially produced the broad appearance of the main sequence termination 
point. 

If we take into account a patchy or clumpy distribution of reddening
towards NGC 2173, we can reproduce the appearance of the termination
point.  The drawback is that the red clump is also affected, resulting
with a too broad color range, while the observed clump is very well
defined and compact.

In addition to a $30 \%$ percentage of binaries we assumed that the star 
formation process of this cluster persisted during an extended age interval,
instead of being instantaneous as usually assumed in star clusters.
In this case the blue envelope of
the main sequence determines the final age of the star formation
process with accuracy, while the initial age is well defined by the MS
red edge. However the presence of binaries might affect the shape of the MS
red edge and make more difficult the determination of the initial age. From a
number of simulations we derived a value of about $ 3 \times 10^8$
years for the age dispersion able to reproduce the termination point
feature (the right panel of Figure 1 shows such a simulation with Z=0.0022).
This dispersion in initial age produces a significant
dispersion in color for the MS, as evident in the simulations, only if
the cluster metallicity Z is greater than about 0.0015.  In fact,
varying the age for lower metallicities the position of the track B
point (beginning of the overall contraction phase at the central
H-exhaustion) moves vertically, increasing the luminosity without
changing its color.
This hypothesis of an extended period of star formation in NGC 2173 meets
some problems on the ground of the theory of cluster formation.
We will discuss a few points related to this possibility in 
section 5.4 

In the following subsection the comparison method between simulations and
data will be described in detail.

\subsection{Global criterium}

Our preliminary simulations show that the chemical composition of this cluster 
is reasonably in the Z range between 0.001 and 0.004, which is inside
the metallicity range reported by observations. 
In this range of metallicity,  the slope of the reddening line 
($A_V/E(V-R)=4.17$ adopting $A_V/E(B-V)=3.1)$ is nearly the same as that of
the line connecting the red clump position in the synthetic CMD at varying 
metallicity. Consequently the overlap of
the synthetic  to the observed clump defines at the same time both the 
distance modulus and the reddening for each
chemical composition. In the following the leading role is played by 
the clump.

\noindent
{\bf First step:} The distance modulus and the  reddening values 
are defined by the overlap of the data with a series of synthetic CMDs 
with a range of metallicities  
and ages suggested by a preliminary analysis with isochrones. 
In particular attention is paid  to the fitting of the color and 
luminosity of the clump.

Table 1 shows the results of the analysis.

\begin{table}
\caption[2]{NGC 2173 Global Criterium: Parameters}
%\begin{scriptsize}
\begin{center}
\begin{tabular}{cccccccccc}
\tableline
\noalign{\smallskip}
%\tableline
 	    Z  &  $\overline{DM}$ &  $\overline{E(V-R)}$ &  $\overline{E(B-V)}$\\
\noalign{\smallskip}
\tableline

	 0.0015&   18.46 &         0.084 &           0.113\\
         0.0020 &   18.46 &        0.076  &          0.103 \\
%         0.0025&   18.46 &        0.066 &           0.088 \\
\noalign{\smallskip}
\tableline
\end{tabular}
\end{center}
%\end{scriptsize}
\label{t1}
\end{table}

We estimate that the maximum uncertainty in color is of the order of 
$\pm 0.008$ and in V magnitude  is 
of the order of $\pm 0.08$, \footnote{This is the combined uncertainty due to
the photometric calibration,  the reddening
and the distance modulus} so that the 
sides of the "uncertainty box" 
are $\Delta c$=0.016  and $\Delta v$=0.16 respectively. 
For the  metallicity range considered we obtain a distance modulus 
of 18.46 mag. This value is adopted for the NGC 2173 analysis.  

\noindent
{\bf Second step:} The detailed comparison with the simulations is performed
as follows:

$\bullet$ N synthetic diagrams are computed for a fixed age (and age interval
in the case of NGC 2173), metallicity, binary percentage  and 
IMF slope with the same number of stars on the main sequence as in the 
observed CMD. The models only differ in the random seed used for the generation 
of the synthetic CMD.

$\bullet$ An exploration of the uncertainty box is made by moving the 
zero point of each of the N synthetic diagrams on the predefined nodes 
of a grid inside the "uncertainty box". 
For each node of the grid the quantities $f1$ms, $f2$ms
and $f1$red are computed.

$\bullet$ After the exploration of all the nodes inside the "uncertainty 
box" the minima of the functions  ($f1ms_{min}$, $f2ms_{min}$, 
$f1red_{min}$)
are stored together with the correspondent displacement $dv$ and 
$dc$.\footnote{
The generation of the synthetic CMDs is done N times  to suppress statistical
fluctuations of stochastic nature in the Monte Carlo simulations. The average 
value of the fitting function minima is in fact more significant than a single
determination, even if lower values of $f1ms_{min}$ and  $f2ms_{min}$
may exist for the given setup (they are less representative).}

$\bullet$ At the end of the N simulations the average values of the functions 
minima
($\overline{f1}ms_{min}$, $\overline{f2}ms_{min}$ and $\overline{f1}red_{min}$)
are computed in addition to their variances and mean values of the 
displacement of the zero point ($\overline{dv}$ and $\overline{dc}$).

Table 2 displays the average values of the fitting functions for each tested 
metallicity and for various ages with the previously defined dispersion.
{\it We assume  that the functions minima correspond to the age values
that best reproduce the data}. The related variances allow us to evaluate 
the uncertainty on the determined age.

\begin{table}
\caption[2]{NGC 2173 Global Criterium: fitting functions mean
values and variances}
%\begin{scriptsize}
\begin{center}
\begin{tabular}{cccccccccc}
\tableline
\noalign{\smallskip}
\noalign{\smallskip}
%\tableline
Z  & Age (Gyr)    & $\overline{f1}ms_{min}$ &$ \sigma$ &
 $\overline{f2}ms_{min}$ &$ \sigma$ & $\overline{f1}red_{min}$  &$ \sigma$ &
$\overline{dv}$& $\overline{dc}$  \\
\tableline
\noalign{\smallskip}
\noalign{\smallskip}
0.0016& 1.2-1.5&  8.04 & $\pm$0.79 & 2850 & $\pm$428 & 11.5&  $\pm$1.7 &       0.03 &     0.008 \\

       & 1.3-1.6&  4.86 & $\pm$0.37 & 1780 & $\pm$308 & 6.4 &  $\pm$0.9 &       0.05 &     0.008 \\

       & 1.4-1.7&  3.46 & $\pm$0.39 &1090  & $\pm$257 & 6.8 &  $\pm$0.9 &       0.05 &     0.006 \\

       & 1.5-1.8&  3.74 & $\pm$0.35 &1210  & $\pm$232 & 8.7 &  $\pm$3.3 &       0.05 &     0.006 \\

       & 1.6-1.9&  4.47 & $\pm$0.29 &1510  & $\pm$119 & 9.8 &  $\pm$2.1 &       0.01 &    -0.008 \\

       & 1.7-2.0&  9.03 & $\pm$0.4  &4650  & $\pm$384 & 9.1 &  $\pm$1.5 &       0.03 &    -0.008 \\

 0.0020& 1.2-1.5&  5.94 & $\pm$0.36 &2140  & $\pm$289 & 10.8&  $\pm$1.8 &       0.07  &     0.008 \\

       & 1.3-1.6&  3.94 & $\pm$0.56 &1400  & $\pm$262 & 6.8 &  $\pm$1.2 &       0.07 &     0.008 \\

       & 1.4-1.7&  3.34 & $\pm$0.47 &1110  & $\pm$251 & 7.8 &  $\pm$1.4 &       0.07 &   -0.002   \\

       & 1.5-1.8&  3.64 & $\pm$0.56 &1150  & $\pm$150 & 9.2 &  $\pm$1.5 &       0.05 &    -0.008 \\

       & 1.6-1.9&  7.90 & $\pm$0.78 &3777  & $\pm$486 & 8.9 &  $\pm$1.5 &       0.06 &    -0.008 \\

       & 1.7-2.0&  15.40& $\pm$0.87 &8350  & $\pm$698 &11.6 &  $\pm$1.9 &       0.08 &    -0.008 \\

 0.0022& 1.2-1.5&  5.92 & $\pm$0.36 &2110  & $\pm$296 &12.3 &  $\pm$1.8 &       0.08 &     0.008 \\

       & 1.3-1.6&  4.13 & $\pm$0.63 &1650  & $\pm$434 & 7.1 &  $\pm$1.8 &       0.08 &     0.006 \\

       & 1.4-1.7&  3.38 & $\pm$0.36 &1150  & $\pm$316 & 9.1 &  $\pm$2.3 &       0.08 &     -0.004 \\

       & 1.5-1.8&  4.23 & $\pm$0.47 &1570  & $\pm$316 &10.8 &  $\pm$1.5 &       0.06 &    -0.008  \\

       & 1.6-1.9&  7.90 & $\pm$0.78 &3777  & $\pm$486 & 8.9 &  $\pm$1.5 &       0.06 &    -0.008 \\

       & 1.7-2.0& 15.40 & $\pm$0.87 &8350  & $\pm$698 & 11.6&  $\pm$1.9 &       0.08 &    -0.008  \\
\noalign{\smallskip}
\tableline
\end{tabular}
\end{center}
%\end{scriptsize}
\label{t1}
\end{table}

We point out that $f1red$ values equal to or greater than 10 correspond to 
a simulated clump either only marginally overlapping, or separated from 
the observed one. So only solutions with average $f1red$ lower than 10 
will be considered.

For all the metallicities of Table 2 both $\overline{f1}ms_{min}$ and
 $\overline{f2}ms_{min}$ present a minimum in correspondence 
of the SF beginning 1.7 Gyr ago and stopping  1.4 Gyr ago. 

Moreover  the relative mean displacements 
$\overline{dv}$ and
$\overline{dc}$ coincide. At the same time the value of $\overline{f1}red_{min}$
 is lower than 10.

In the analysis we give  more weight to the main sequence
than to the red region of the CMD due to the possible presence of errors 
in the conversion from theoretical to observational quantities, which should
be more pronounced for the red giants at lower effective temperatures. For this 
reason we do 
not require the minimization of the  function $\overline{f1}red_{min}$ in 
correspondence of the acceptable solution, but simply a  value lower than 10.

For
metallicities higher than 0.0022 it is not possible to fit simultaneously
the MS and the red clump stars, as for increasing metallicity the color 
separation between turn-off and red clump stars decreases.

The conclusion  is that the MS and the red clump stars can be fitted at the 
same time (within the tolerance imposed by the "uncertainty box") only for
values of the metallicity comprised inside the range 
$ 0.0016 < Z < 0.0022$ and for a  constant SFR between  1.7 and 1.4  
Gyr ago.

 From the data of Table 2 it is also possible 
to determine graphically the uncertainty on the age as a function of the
metallicity, as described in section 5.2. A rough determination indicates 
an uncertainty of the order of $\pm0.15$ Gyr independent of metallicity.

All the simulations take into account the presence of binaries with a 
percentage of $30 \% $. The best reproduction of the 
feature of the cluster MS termination point is with higher metallicities
in the range given in Table 2.
     
Figure 1 shows the observed CMD (right panel) and a simulation for
the metallicity Z = 0.0022 (left panel) with a SFR constant in the age interval
1.4-1.7 Gyr.  

A shallower IMF slope ($\alpha$ =1.35) does not produce significant changes
in the previously reported results, as the involved range of masses is not large
enough to cause significant variations of the stellar distribution in the CMD.

\subsection {Test of the method}

To test the reliability of the criteria adopted in evaluating the results
we computed a synthetic diagram with the characteristics assigned to NGC 2173.
This simulated CMD is considered as an observed one. To this CMD we applied
our method looking for the values of age and metallicity best reproducing
its stellar distribution.

The adopted values for this test were : time interval for the prolonged star
formation between 1.5 and 1.8 Gyr, metallicity Z=0.0018, distance modulus
18.5 mag, reddening E(B-V)=0.056, binary percentage 30$\%$ and IMF slope
2.35.
To simulate observational errors we introduced a random displacement of the 
cluster CMD zero point with $dv = -0.02$ and $dc = 0.006$.

From the {\bf first step} of the analysis we obtained an average DM = 18.47 
together with the metallicities Z = 0.0015 and 0.0020. The average 
reddening is  E(B-V) = 0.076 and 0.066 respectively.
In the {\bf second step} the functions $f1$ms, $f2$ms and $f1$red are derived 
for the metallicities 0.0016, 0.0020 and 0.0022. 
Figure 2 shows the trend of the
average $f1ms_{min}$ (solid line) as a function of the {\bf final age} of the 
star formation process for those metallicities (the period of prolonged
and constant SF adopted in the simulations lasts 0.3 Gyr, as in the "fake
cluster").
The dashed lines represent the average $f1ms_{min} \pm 1 \sigma$.
The uncertainty on age is evaluated from the intersection of the straight line, 
(traced parallel to the time axis) and passing through the minimum of the upper
dashed line corresponding to $f1ms_{min} + 1\sigma$ with the lower dashed
line relative to the average $f1ms_{min} - 1\sigma$. In Figure 2 the thicker
 vertical
bars identify the interval of age corresponding to the evaluated uncertainty.

The final age of the prolonged star formation determined from this test 
analysis is 1.5 Gyr for the two lower
metallicities and 1.4 Gyr for Z=0.0022.
Considering altogether the uncertainties from the three chemical compositions
we estimate that the {\bf final age} is 1.5 $\pm$ 0.15 Gyr.
We have repeated the test with the same "fake cluster" adopting in the
simulations a single age instead of a constant SF during an interval of
0.3 Gyr. The minimum of the mean fitting functions determines age values
equal to 1.5 or 1.6 Gyr (depending on the chemical composition adopted).
However the values of the two fitting functions at the minimum are 
approximately a factor of two greater than in the case of the analysis
with the prolonged star formation, clearly suggesting that the solution with
a continuous SF during 0.3 Gyr must be preferred.  

The conclusion is that this method can recover the initial parameters
with an uncertainty strictly limited to a well defined "confidence box",
as introduced in section 4.
The minima of the fitting functions
for the test of the method in the case of prolonged SF are three times lower 
than those relative to
the observed NGC 2173 CMD, even if the trend of the functions is almost the
same. This difference can be ascribed to the peculiar features  of the cluster
described at the beginning of this section  and to the field
subtraction, together with many other small uncertainties in the assumptions,
all summed up: the models have all exactly the same assumptions and likely
don't include a complete description of {\it all} the processes that go on 
in real stars and stellar systems.

\subsection {Partial criterium}  
 
The second criterium requires that the main sequence and the luminosity of 
the clump, but not its color, are reproduced  correctly.
 In the considered metallicity range the difference in
luminosity between the MS termination point and the red clump depends 
significantly on the age, and not very much on Z.

Distance modulus and  reddening are obtained from a comparison of
the data with a series of simulations for ages and various chemical abundances 
in the range determined for the cluster. The best fit is obtained from the 
constraint that the synthetic diagram ought to fit both the MS
termination point and the clump luminosity. The results are reported 
in Table 3.

\begin{table}
\caption[2]{NGC 2173 Partial Criterium: parameters}
%\begin{scriptsize}
\begin{center}
\begin{tabular}{cccccccccc}
\tableline
\noalign{\smallskip}
%\tableline
            Z  &  $\overline{DM}$ &  $\overline{E(V-R)}$ &  $\overline{E(B-V)}$\\
\noalign{\smallskip}
\tableline
%              0.0025&   18.539&    0.049&         0.066\\
              0.0030&   18.50 &    0.041&         0.056\\   
              0.0040&    18.48 &    0.028&         0.037\\
\noalign{\smallskip}
\tableline
\end{tabular}
\end{center}
%\end{scriptsize}
\label{t1}
\end{table}

For each metallicity in Table 3 a range of age was explored in a similar way 
to the global analysis producing the results listed in Table 4.

\begin{table}
\caption[2]{NGC 2173 Partial Criterium: fitting functions mean values and
variances}
%\begin{scriptsize}
\begin{center}
\begin{tabular}{cccccccccc}
\tableline
\noalign{\smallskip}
%\tableline
Z  & Age (Gyr)    & $\overline{f1}ms_{min}$ &$ \sigma$ &
 $\overline{f2}ms_{min}$ &$ \sigma$ &$\overline{dv}$&$\overline{dc}$  \\
\tableline
% 0.0025& 1.3-1.6 & 6.13 & $\pm$0.42 &  2190 & $\pm$290  &      0.035 &   0.0075  \\
%
%      & 1.4-1.7 & 3.55 & $\pm$0.52 &  1310 & $\pm$272  &      0.05  &   0.0075  \\
%
%      & 1.5-1.8 & 3.16 & $\pm$0.38 &   979 & $\pm$175  &      0.05  &  -0.002   \\
%     
%      &  1.6-1.9& 3.89 & $\pm$0.45 &   1380& $\pm$141  &     -0.02  &  -0.0075  \\
%
0.0030&  1.3-1.6& 4.91 & $\pm$0.62 &   1970& $\pm$293  &     0.08  &  0.008   \\

      &  1.4-1.7& 3.65 & $\pm$0.46 &   1500& $\pm$226  &      0.07 &   0.002 \\
 
      &  1.5-1.8& 3.03 & $\pm$0.34 &   1170& $\pm$357  &       0.06&   -0.008 \\

      &  1.6-1.9& 6.29 & $\pm$0.63 &   2580& $\pm$471  &       0.06&   -0.008 \\

0.0040&  1.3-1.6& 5.90 & $\pm$0.35 &   2080& $\pm$309  &       0.08&    0.0    \\

      &  1.4-1.7& 3.53 & $\pm$0.41 &    1670& $\pm$171 &        0.08&   -0.008 \\

      &  1.5-1.8& 5.91 & $\pm$0.43 &    2730& $\pm$360 &   0.08     &  -0.008 \\
\noalign{\smallskip}
\tableline
\end{tabular}
\end{center}
%\end{scriptsize}
\label{t2}
\end{table}

Solutions at higher metallicities than in the global case
are obtained as the partial criterium does not constrain  the clump color. 
For Z=0.0030 the solution for the cluster final age is 1.5 Gyr
with an uncertainty of $\pm 0.15$ Gyr, while for Z=0.004 the minimum
corresponds to a slightly younger final age, 1.4 Gyr. The solution for the 
metallicity Z=0.004 involves a too low value of
the reddening. 

Obviously these cases, that do not reproduce the clump
color, produce too high values of the function $f1red$ (always
higher than 10 and not reported in Table 4).
Table 5 summarizes the accepted solutions. Figure 3 shows a simulation
for the   case  Z=0.003 (partial criterium) with the input  values 
of Table 3 and the age indicated by Table 4. 

\begin{table}
\caption[2]{NGC 2173: summary of results}
%\begin{scriptsize}
\begin{center}
\begin{tabular}{cccccccccc}
\tableline
\noalign{\smallskip}
%\tableline

                  &     Z  & DM & E(B-V) &  Age (Gyr)& dv& dc \\
%\tableline
\noalign{\smallskip}
\tableline
  global criterium&   0.0016&  18.46&  0.111 & 1.4-1.7 &  0.05 &   0.006 \\

         "        &   0.0020&  18.46&  0.103 & 1.4-1.7 &  0.07 &  -0.002 \\         

          "       &   0.0022&  18.46&  0.097 & 1.4-1.7 &  0.08 &  -0.004 \\              

%  partial criterium&  0.0025& 18.54& 0.066  & 1.5-1.8 &  0.05 &  -0.002 \\
%
   partial criterium&  0.0030& 18.50& 0.056  & 1.5-1.8 &  0.06 &  -0.008 \\

         "         &  0.0040& 18.48& 0.028  & 1.4-1.7 &  0.08 &  -0.008 \\
\noalign{\smallskip}
\tableline
\end{tabular}
\end{center}
%\end{scriptsize}
\label{t1}
\end{table}

\subsection{Inadequate field subtraction or prolonged star formation?}

One could argue that some of the NGC 2173 features could have their origin
in an imperfect subtraction of the LMC background field, which is described
in detail in Paper I of this series. The process is illustrated in Figure 3 
of that paper and it can be judged as very successful from the great
similitude between the 'field' and 'subtracted stars' panels in that figure,
although possible gradients in the background composition cannot be ruled
out. One must note, however, the following: in the magnitude range $19.5
\le V \le 20.75$, which comprises both the main sequence turnoff and the
subgiant branch region, there are few stars in the field, and therefore
the field subtraction will have little effect on the morphology and stellar
distribution of these regions. 
The field subtraction may have some influence on the width of the main sequence
(point 3 at the beginning of section 5): in Figure 3 of Paper I  the cluster 
CMD is shown before and
after the field subtraction, and a change in the width of the main sequence
is apparent.
  
In an attempt to estimate the effect of stochastic fluctuations in the
field CMD on our procedure to statistically remove field stars from
the cluster CMD, we performed the following experiment.  We computed a
synthetic CMD for a fixed age (1.5 Gyr), metallicity (Z=0.0022) and
percentage of binaries (30$\%$). The total number of stars in this
synthetic CMD was fixed by the criterion of having the same MS stars,
around the turnoff region, as the observed cluster CMD. Figure 4a ( left panel)
shows this simulation with fixed age, and the comparison with the
right panel of Figure 1 clearly reveals the effect of a prolonged star
formation in the latter. We then randomly split the observed field CMD
in two parts, having the same area.

One of them was added to the simulated CMD (Fig. 4a, right panel ) and the 
other one
was considered as control "field" population (Fig. 4b, left panel), 
and used to
statistically decontaminate the "simulated+field" CMD. The experiment was
repeated 5 times, in order to observe the effect of different random 
extraction of the two field sub-populations (the one to be added to
the simulated CMD and the one to be used as control field). 
Fig. 4b (right panel)  shows a typical result. This experiment demonstrated
that, as expected, only the regions of the CMD that are sparsely
populated (e.g., the SGB and the  RGB) can be affected by
statistical fluctuations between the field population that we used as
"control field" and the one that is actually contaminating the cluster
CMD. On the contrary, the shape of the turnoff region represents a "robust"
characteristic of the cluster and is preserved after the subtraction.

In conclusion, it is evident that the field subtraction
procedure does not affect the "single-age" appearance of the turnoff
region of the single-age simulated CMD, and hence the suggestion of an 
age spread required to reproduce the 
NGC 2173 CMD is not the consequence of a poor field subtraction.

For the LMC cluster NGC 1850 Vallenari et al. (1994) suggested a prolonged
star formation activity, since that cluster shows features pointing to the
presence of an 8 Myr population and an older one of about 70 Myr
(this age spread is shorter than the supposed one for NGC 2173).
   
 However the  hypothesis that in a cluster the star formation is going
on during a period of the order of 0.3 Gyr is hardly supported from
the theory  of the formation and evolution of 
star clusters (see the review by Kroupa, 2001).
    
From a comparison between CO clouds and clusters in the LMC Fukui et al.
(1999) and Yamaguchi et al. (2001) suggest that stellar clusters are actively 
formed over $~50 \%$ of a cloud lifetime  as about half of the CO clouds are 
associated with the youngest stellar clusters. 
They estimate the evolutionary time scale of the Giant Molecular Clouds (GMCs)
as follows: GMCs form stars in a few Myr after their birth, form clusters 
during the next few Myr and dissipate rapidly due to stellar UV photons in the
subsequent few Myr. 
At the present day our Galaxy forms only unbound open clusters, whereas
the LMC is forming populous clusters, that are perhaps gravitationally 
bound and resemble globular clusters in the Galactic halo, although the
number of stars in a populous cluster is about 10 times smaller than in a 
globular cluster. The gravitational field of the LMC is perhaps weaker and
less flattened than that of the Galaxy, favoring the formation of more
massive fragments that may lead to form populous clusters. 
This fact points out that environmental differences can give rise to differences
for the cluster formation in the Galaxy and the LMC, even if it is unlikely 
that star formation in a cluster can last for a time as long as suggested
by our simulations for NGC 2173.

Binary clusters or mergers might be considered as a possible explanation
for the NGC 2173 CMD. As far as binary clusters are concerned,
Fujimoto \& Kumai (1997) suggested that the components of a binary cluster 
form together, as a result of oblique cloud-cloud collisions,
 being coeval or having small age differences compatible with
cluster formation time scale. 
A statistical study of binary and multiple clusters in the LMC by Dieball et 
al. (2002) finds that multiple clusters are predominantly young. Old groups
or groups with large internal age differences are mainly located in the densely
populated bar region, thus they can be easily explained as chance 
superpositions. 

On the other hand also the probability of close encounters between star 
clusters leading to a tidal capture is considered to be relatively small or
even very unlikely. As matters stand the only alternative to a prolonged star
formation seems to be that of a merging of two clusters with similar 
metallicity and not much different ages (of the order of 0.3 Gyr).  

\section{SL 556}

SL~556 (Hodge~4) is a relatively less studied cluster.  $U,B,V$ 
photometry
has been published by Mateo \& Hodge (1986) who estimated a metallicity
of [Fe/H]=-0.7 using various photometric indexes. Olzewski et al. (1991)
determined [Fe/H]=-0.15.  A near infrared RGB slope based abundance of 
[Fe/H]= -0.17 $\pm$ 0.04 for SL 556 (Hodge 4) has been derived by
Sarajedini et al. (2002). These values are higher than other evaluations from
photometric data. Photometric studies have
been published by Sarajedini (1998) and Rich et al. (2001). They analysed
the same HST data, adopted a metallicity [M/H]=-0.68 and found 
an age between 2 and 2.5 Gyr.

SL 556 is a cluster with a significant percentage of binaries,  clearly
visible from a double main sequence. The presence of the binaries
reproduces entirely the shape of SL~556 termination point without
requiring a prolonged SF as in the case of NGC~2173. 
In the analysis we adopted a  percentage of binaries amounting to $40 \%$
with a flat distribution of the mass ratio in the range $0.4 \leq  M_2 / M_1 
\leq 1.0$.

The results from the two criteria are:

1)  Global criterium. Based on isochrones evaluations, we defined a reasonable
range of age (1.8 - 2.1 Gyr)and metallicity (0.003 $\leq$ Z $\leq$ 0.004).
In this range several synthetic diagrams are computed to look for the best
 fit of the clump. The average
values of metallicity, distance modulus and reddening derived from these
simulations are listed in Table 6.

\begin{table}
\caption[2]{SL 556 Global Criterium: parameters}
%\begin{scriptsize}
\begin{center}
\begin{tabular}{cccccccccc}
\tableline
\noalign{\smallskip}
%\tableline
            Z  &  $\overline{DM}$ &  $\overline{E(V-R)}$ &  $\overline{E(B-V)}$\\
\noalign{\smallskip}
\tableline
              
                0.003&   18.43&   0.045&       0.061\\
		0.0035&  18.44&   0.038&       0.052\\
                0.004 &  18.46&   0.021&       0.029\\
\noalign{\smallskip}
\tableline
\end{tabular}
\end{center}
%\end{scriptsize}
\label{t1}
\end{table}

Adopting 18.44 for the distance modulus, for the metallicity and reddening as
given in Table 6, we derive the 
average values for the minima of the functions ( $\overline{f1}ms_{min}$ and 
$\overline{f2}ms_{min}$) varying the 
cluster age. In Table 7 the results of the global analysis are listed.    
{\it For all the considered metallicities the minimum of the MS functions is
found at the age 2.0 Gyr}. 

\begin{table}
\caption[2]{SL 556 Global Criterium: fitting functions mean values and
variances}
%\begin{scriptsize}
\begin{center}
\begin{tabular}{cccccccccc}
\tableline
\noalign{\smallskip}
\noalign{\smallskip}
%\tableline
Z  & Age (Gyr)    & $\overline{f1}ms_{min}$ &$ \sigma$ &
 $\overline{f2}ms_{min}$ &$ \sigma$ & $\overline{f1}red_{min}$  &$ \sigma$ &
$\overline{dv}$& $\overline{dc}$  \\
\tableline

 0.0030& 1.8& 5.06& $\pm$0.51&  2400& $\pm$270& 4.21&  $\pm$0.74&   0.08&     0.008\\

       & 1.9& 2.55& $\pm$0.33&  1080& $\pm$198& 6.73&  $\pm$1.41&   0.07&     0.007\\ 

       & 2.0& 2.38& $\pm$0.31&  1010& $\pm$229& 2.26&  $\pm$0.87&  -0.02&    -0.003\\

       & 2.1& 3.00& $\pm$0.24&  1280& $\pm$181& 2.02&  $\pm$0.44&  -0.08&    -0.008\\

 0.0035& 1.8& 3.96& $\pm$0.40&  1990& $\pm$349& 5.93&  $\pm$0.97&   0.08&    0.008 \\

       & 1.9& 2.40& $\pm$0.39&  1080& $\pm$282& 3.99&  $\pm$1.62&   0.05&    0.001\\ 

       & 2.0& 2.25& $\pm$0.35&   917& $\pm$165& 1.63&  $\pm$0.65&   0.00&    -0.008\\

       & 2.1& 4.17& $\pm$0.38&  1820& $\pm$301& 2.33&  $\pm$0.74&  -0.07&   -0.008\\

 0.0040& 1.8& 5.20& $\pm$0.38&  2470& $\pm$220& 3.00&  $\pm$0.79&   0.08&    0.008\\ 

       & 1.9& 2.21& $\pm$0.33&   942& $\pm$196& 2.80&  $\pm$0.88&   0.08&    0.008\\

       & 2.0& 2.06& $\pm$0.17&   798& $\pm$155& 2.26&  $\pm$1.07&  -0.02&   -0.002\\

       & 2.1& 2.84& $\pm$0.44&  1160& $\pm$248& 3.74&  $\pm$1.15&  -0.08&   -0.008\\   
\noalign{\smallskip}
\tableline
\end{tabular}
\end{center}
%\end{scriptsize}
\label{t1}
\end{table}

Also for the red stars the 
function values give a good fit of the clump. The derived uncertainty
in age is of the same order as for NGC 2173 ($\pm 0.1 Gyr$). 
Figure 5 displays SL 556 data and  a simulation for
Z=0.004 with  parameter values from Table 6 and a  2 Gyr age as
obtained from Table 7.

2) Partial criterium. 
We search solutions without the constraint of fitting exactly the color 
of the clump. From synthetic diagrams we derive  the values of SL 556 distance
modulus and reddening that are  listed in Table 8.

\begin{table}
\caption[2]{SL 556 Partial Criterium: fitting functions mean values and
variances}
%\begin{scriptsize}
\begin{center}
\begin{tabular}{cccccccccc}
\tableline
\noalign{\smallskip}
%\tableline
            Z  &  $\overline{DM}$ &  $\overline{E(V-R)}$ &  $\overline{E(B-V)}$\\
\noalign{\smallskip}
%\tableline
                0.0020 &  18.32&   0.078&       0.105\\
                0.0025&  18.31&   0.068&       0.091\\
\noalign{\smallskip}
\tableline
\end{tabular}
\end{center}
%\end{scriptsize}
\label{t1}
\end{table}

For each of the metallicities in Table 8 we explored a range of ages.  Table 
9 shows
the corresponding average values of the minima of the  functions.
We note that the minima for Z=0.0020 present significantly  higher values 
than for the  other metallicities. This solution is therefore discarded, 
since it corresponds to a worse match of the observed CMD.
The best fit for Z=0.0025 determines an age of
 2 Gyr like that obtained with the global criterium.

\begin{table}
\caption[2]{SL 556 Partial Criterium: fitting functions mean values and
variances}
%\begin{scriptsize}
\begin{center}
\begin{tabular}{cccccccccc}
\tableline
\noalign{\smallskip}
\noalign{\smallskip}
%\tableline
Z  & Age (Gyr)    & $\overline{f1}ms_{min}$ &$ \sigma$ &
 $\overline{f2}ms_{min}$ &$ \sigma$ &$\overline{dv}$&$\overline{dc}$  \\
\tableline

 0.0020& 1.8 &   7.33 & $\pm$0.61&   3530&  $\pm$456&    0.08&    0.008\\

       & 1.9 &  4.66 & $\pm$0.60 &  2210 & $\pm$381 &   0.08 &   0.007\\

       & 2.0 &  3.85 & $\pm$0.39 &  1760 & $\pm$231 &   0.06 &  -0.002\\

       & 2.1 &  3.84 & $\pm$0.33 &  1720 & $\pm$260 &  -0.01 &  -0.008\\

 0.0025& 1.8 &  5.78 & $\pm$0.57 &  2870 & $\pm$269 &   0.08 &   0.007\\  

       & 1.9 &  3.80 & $\pm$0.40 &  1950 & $\pm$209 &   0.08 &  -0.003\\

       & 2.0 &  2.84 & $\pm$0.37 &  1310 & $\pm$220 &   0.06 &  -0.008\\

       & 2.1 &  5.29 & $\pm$0.42 &  2590 & $\pm$227 &   0.02 &  -0.008\\
\noalign{\smallskip}
\tableline
\end{tabular}
\end{center}
%\end{scriptsize}
\label{t1}
\end{table}

Table 10 shows the solutions depending on metallicity for both criteria.

\begin{table}
\caption[2]{SL 556: summary of results}
%\begin{scriptsize}
\begin{center}
\begin{tabular}{cccccccccc}
\tableline
\noalign{\smallskip}
%\tableline

               &     Z  &    DM &   E(B-V) &  Age (Gyr)&          dv  &     dc\\
%\tableline
\noalign{\smallskip}
\tableline
   partial criterium&  0.0025& 18.32&   0.091&    2.0 &  0.060&   -0.008\\

  global criterium  & 0.0030&  18.44&   0.061&    2.0&  -0.020&   -0.003\\
          "         & 0.0035& 18.44 &  0.052 &   2.0 & -0.002 &  -0.008 \\
          "         & 0.0040& 18.44 &  0.029 &   2.0 & -0.020 &  -0.002 \\
\noalign{\smallskip}
\tableline
\end{tabular}
\end{center}
%\end{scriptsize}
\label{t1}
\end{table}

\section{NGC 2155}

NGC~2155 has been studied by Bica, Dottori and Pastoriza (1986)
who estimated a metallicity of [Fe/H]=--1.2. More recently, a
significantly higher value ([Fe/H]=--0.55) has been found by Olszewski
et al. (1991) who performed Ca II triplet spectroscopy of three cluster
giants. 
A more recent determination of the
cluster metallicity from the RGB slope comes from the HST photometry
published by Sarajedini (1998), who estimates [M/H]=-1.1, and
an age of $\sim 4$ Gyr. A renewed analysis (of the same HST data used by 
Sarajedini) by Rich et al. (2001) gave a global metallicity of
[M/H]=-0.68 and an age of 3.2 Gyr.
Piatti et al. (2002) derive a metallicity [Fe/H]=-0.9 from Washington
photometry and an age of $3.6 \pm 0.7$ Gyr.

The cluster data in Figure 7 (left panel) show a peculiar feature of NGC 2155:
the presence of a gap at $V \sim 21$ and $V-R \sim 0.25$.
This  separates the MS termination point from a stage of relatively
slow evolution towards the red part of the CMD, in which H-burning occurs
in a shell around an almost isothermal core.

A simple comparison with isochrones suggests an age of the order of 3 Gyr.
So more evolved cluster stars have masses in the range $1.2 - 1.3
M_{\odot}$. In this mass range there are a few open questions on  
treatment and efficiency of core overshoot (Aparicio et al., 1990 discussed
this problem in details, Chiosi 1999, Bertelli 2000).    

Features depending critically on the presence and efficiency of overshoot,
in addition to its treatment are: the presence or absence of the gap at the
MS termination point and its width, the location in the HR diagram of the
beginning of the slow phase toward the red during shell H-burning, the stellar
distribution in the CMD during this phase and the number ratio between red 
giants and  MS stars. 

As in Bertelli et al. (1994) and in Girardi et al.
(2000), the function FLUM is defined as the indefinite integral of the 
initial mass function by number of stars (equation (1)). 
The difference between  any two values 
of FLUM is proportional to the number of stars located in the corresponding 
mass interval. A rapid flattening of FLUM in a magnitude interval
implies the presence of a gap in the CMD.
Figure 6 (left panel) displays the function FLUM  versus the visual
magnitude for two isochrones (at the LMC distance, age 3 Gyr and Z=0.001)
for two different treatments of core convective mixing. 
The left panel of Figure 6 shows the $Y^2$ isochrone at 3 Gyr (dashed line).
Evolutionary models by Yi et al. (2001) are computed without overshoot for
ages $\geq$ 3 Gyr. The solid line shows the 3 Gyr isochrone
from Padova models with moderate overshoot according to the scheme described 
in Girardi et al. (2000). 

In Figure 6 (right panel) the functions FLUM are plotted from isochrones 
with Z=0.004. Models without overshoot produce FLUM functions without
gap for the  metallicity range considered. The Padova models
show an FLUM function with a gap, corresponding to a faster phase of the
evolution (lower number of stars in the relative CMD bins).  
From Figure 6  we point out that the gap, centered at $V = 21.25$, broadens 
for increasing metallicity in the range $0.001 \leq Z \leq 0.004$
with the Padova models. The reason of the gap widening for increasing 
metallicity is related to the increase of radiative opacities, which causes
larger convective cores and also larger overshooting regions.

To reproduce the observed gap we must use models with convective overshoot.
There are also two contrasting effects to be taken into account: 
the gap increases with metallicity and the reddening value decreases
with metallicity. There is a low boundary in metallicity ($Z=0.002$), below 
which the gap is too small in comparison to the observed one, and an upper 
limit ($Z=0.003$) above which the reddening becomes too low.

The results in this tightly constrained metallicity range  are given in
Table 11. 

\begin{table}
\caption[2]{NGC 2155 Global Criterium: parameters }
%\begin{scriptsize}
\begin{center}
\begin{tabular}{cccccccccc}
\tableline
\noalign{\smallskip}
\noalign{\smallskip}
%\tableline
            Z  &  $\overline{DM}$ &  $\overline{E(V-R)}$ &  $\overline{E(B-V)}$\\
\noalign{\smallskip}
\tableline
              0.0020&   18.36&    0.021&         0.028\\
	      0.0025&   18.36&    0.015&         0.020\\
              0.0030&   18.36&    0.011&         0.015\\
\noalign{\smallskip}
\tableline
\end{tabular}
\end{center}
%\end{scriptsize}
\label{t1}
\end{table}

For NGC 2155 we used only the global criterium because of the metallicity
constraints. The partial criterium would require metallicities out of the
allowed range.

\begin{table}
\caption[2]{NGC 2155 Global Criterium: fitting functions mean values
and variances}
%\begin{scriptsize}
\begin{center}
\begin{tabular}{cccccccccc}
\tableline
\noalign{\smallskip}
Z  & Age (Gyr)    & $\overline{f1}ms_{min}$ &$ \sigma$ &
 $\overline{f2}ms_{min}$ &$ \sigma$ & $\overline{f1}red_{min}$  &$ \sigma$ &
dv& dc  \\
\tableline

 0.0020& 2.7& 7.39& $\pm$0.73&  1780& $\pm$234&    3.68&   0.74&        0.08&     0.008\\
       & 2.8& 6.56& $\pm$0.70&  1260& $\pm$154&    3.89&   0.51&        0.06&     0.007\\
       & 2.9& 6.74& $\pm$0.97&  1360& $\pm$202&    3.85&   0.43&        0.05&     0.006\\
       & 3.0& 7.32& $\pm$0.67&  1720& $\pm$277&    4.20&   0.71&        0.02&     0.000   \\

 0.0025& 2.7& 4.91& $\pm$0.60&   749& $\pm$189&    3.51&   0.45&        0.07&     0.004\\
       & 2.8& 4.62& $\pm$0.38&   762& $\pm$112&    3.18&   0.56&        0.07&    -0.004\\
       & 2.9& 4.73& $\pm$0.46&   914& $\pm$172&    3.79&   0.67&        0.05&    -0.005\\
       & 3.0& 5.32& $\pm$0.57&  1210& $\pm$184&    3.70&   0.52&        0.00  &  -0.007\\

 0.0030& 2.7& 4.21& $\pm$0.83&   680& $\pm$258&    3.17&   0.51&        0.08&    -0.007\\
       & 2.8& 4.13& $\pm$0.66&   625& $\pm$124&    3.41&   0.56&        0.07&    -0.008\\
       & 2.9& 4.65& $\pm$0.48&   808& $\pm$186&    3.58&   0.40&        0.03&    -0.008\\
       & 3.0& 6.16& $\pm$0.57&  1250& $\pm$202&    3.76&   0.69&       -0.08&    -0.008\\
\noalign{\smallskip}
\tableline
\end{tabular}
\end{center}
%\end{scriptsize}
\label{t1}
\end{table}

The solution at Z=0.002 produces too high minimum values of the 
functions $\overline{f1}ms_{min}$ and $\overline{f2}ms_{min}$, so that it 
is not acceptable, see Table 12. Concerning the solution Z=0.0025 the 
minima of the average functions for the main sequence  do not
coincide in age. For the metallicity Z=0.003, there is a minimum at age 
2.8 Gyr for both functions, but for $f1$ it is not well defined. Its value    
is significantly higher than obtained from previous results. The average 
displacement in 
color of the zero point of the synthetic diagram is relative to function $f1$
average value only. This is in contrast to the other two clusters,
where the  displacements were coincident for the two functions. 
One more problem comes from the too low value of the reddening in the
acceptable solutions. These problems might suggest
that a way to overcome this difficulty can be found with a different
overshoot parameterization for  evolutionary models with
$1.2 - 1.3 M_{\odot}$. A more efficient overshoot may
better reproduce NGC 2155 features. 

A further problem in the synthetic CMDs simulating this cluster comes from
red clump stars, as for equal number of MS stars the number of red clump
stars in the synthetic diagrams is about one half of that in the cluster
data. This discrepancy would also be reduced if in the models a higher 
overshoot efficiency were adopted during the main sequence phase. In
addition to this effect a larger convective core would also move the stars 
to the  right in the CMD above the MS termination 
point gap, hence better  reproducing this feature.   

Part of the difference may also be ascribed to the  subtraction process
of the LMC field stars. 

%________________________________________________________________________

\section{SUMMARY}
\label{summ}

Three LMC clusters, NGC 2173, SL 556 and NGC 2155, are analyzed by means of
synthetic CMDs, based on stellar models by  Girardi et al.
(2000). To derive their age and metallicity we evaluate the best fit of
simulations to observations using two fitting functions, 
which contain  suitable  combinations of star  distribution inside the 
bins of a grid covering  the CMD. 
We  vary the parameters
which characterize the synthetic cluster (age, metallicity and IMF)     
looking for those values which minimize the fitting
functions. These parameters represent the solution of the problem.

For the comparison we used two criteria  (described in section 5):

i) a global criterium where all the main characteristics of the cluster must
be fitted (main sequence turn-off and termination point ,  
luminosity and  color of the red clump);

ii) a partial criterium where
 the fitting of clump  color is no more required. 

The significance of the second criterium is that it takes into account
the uncertainty on  clump color depending on the conversion  from
the theoretical to the observed CMD and
on the input physics of the stellar models. From the partial criterium
we obtain solutions for the metallicity different from those of the
global criterium, as allowing that the real clump color may be different
from the considered one the solution fixes a different metallicity for 
a given set of models.

For each cluster we present a set of solutions 
depending on the metallicity and the  adopted criteria.  

NGC 2173. We interpret the unusual shape of the termination point as due to 
a peculiar characteristic of this cluster: an age dispersion 
of about  0.3 Gyr. We have shown that this unusual shape is not likely
related to an imperfect field star subtraction, to differential reddening, or
to the presence of binary stars, as suggested by the many simulations
performed to reproduce its CMD features. 
With the constraints imposed by the global criterium the solutions for 
the metallicity are in the range $0.0016\le  Z
\le 0.0022$ and the correspondent age with an age dispersion in the process
of star formation is from 1.4 to 1.7 Gyr for a distance modulus equal to 18.46.
The partial criterium allows solutions for higher values 
of the metallicity. We derive the age with a prolonged star formation from 1.5
to 1.8 Gyr in connection  
with  Z=0.003 and from 1.4 to 1.7 Gyr for Z=0.004, with distance
modulus respectively 18.50 and 18.48.  
The existence of a star formation going on for a significant period (0.3 Gyr) 
is a very unusual characteristic for a globular cluster,  and a comprehensive 
discussion of the possible origin of such an age dispersion in NGC 2173 and
of its implications exceeds the scope of the current paper.
In the case of $\omega Cen$ (Ferraro et al. 2002 and references therein)
the complex structure of the red giant branch, ascribed to the presence of
multiple populations, points toward a multiple merging and/or accretion event
in the past history of $\omega Cen$ to explain the structural and kinematical
properties of the various cluster subpopulations. 
In a relatively small cluster as NGC 2173 (at least compared to $\omega Cen$)
a period of star formation more extended than the typical timescale of SNII,
which would likely remove any remaining gas, is extremely unlikely, and a 
merger of two clusters with slightly different ages could be a possible 
explanation.

For the other two clusters the features of the main sequence termination point 
are different from those of NGC 2173, so there are no such problems in 
interpreting their CMD.

SL 556. In the range of metallicity 0.003-0.004 the synthetic models 
fit at the same time MS and red clump (global criterium) with the age
2 Gyr and a distance modulus equal to 18.44. 
Its CMD shows a more significant percentage of binaries and in the simulations
we adopted a percentage of $40 \%$. In the contest of the partial
criterium lower metallicities are considered. In correspondence of the
value Z=0.0025 a solution for the age equal to 2 Gyr is  obtained 
 but with a distance modulus equal to 18.32.

NGC 2155. This cluster is characterized by a gap in the distribution 
of the stars at the MS termination point. It is strictly related to the 
treatment and the importance of  convective overshoot 
and to metallicity. Our models impose a lower limit to the 
metallicity (Z=0.002) on the base of the amplitude of the gap (which 
depends on the metallicity) and an upper limit (Z=0.003) requested by a non
zero value of the reddening. In this range we find a solution
satisfying the global criterium with an age equal to 2.8 Gyr $\pm0.15$ Gyr.
Part of the difficulties presented by this study might be ascribed 
to the insufficient amount of overshoot adopted in the stellar 
models. On the other hand the mass interval  concerning the evolved
stars of the cluster (1.2-1.3 $M_{\odot}$) is quite critical as far as
the convective overshoot is concerned (Aparicio et al 1990).

It is important to stress that the age, metallicity and reddening 
found in this paper are proper to the set of stellar models in use, 
and are subject to small systematic errors that derive from several
possible causes. Just to mention a single one: the red clump luminosities
of Girardi et al. (2000) models are about 0.15 mag systematically
fainter than in similar models (Salasnich 2000), due to 
the different input physics adopted by different authors. This means 
that distance measurements based on the global criterium will tend to
be slightly lower (by about 0.15 mag) than the distances one could 
find if more luminous clump models were used. Clearly the differential
values of distance (from cluster to cluster) derived in this work 
are more accurate than the absolute values. Similar remarks apply
to the other cluster parameters we have derived.  

We have been able to find sufficiently precise values
for the age and metallicity of each cluster within plausible ranges of 
reddening and distance modulus. A good match of the different features 
observed in the clusters CMDs with the synthetic CMDs is obtained
 both qualitatively
and quantitatively (as represented in Figures 1, 5 and 7). In particular   
the good match of the shape of the MS termination point indicates 
that the amount of CCO assumed in the models is adequate. Only
in the case of NGC 2155 we find a poorer agreement between models and data,
and we conclude that the amount of CCO assumed in the models for that
age-mass may be insufficient. This may imply that the decrease of $\Lambda_c$
as a function of mass assumed in the models may be too abrupt for this
metallicity. 

%____________________________ ACKNOWLEDGEMENTS _____________________

\acknowledgements

We would like to thank Yuen Ng for a careful reading of the manuscript
and for his suggestions, helping to improve the paper.
The referee's remarks and comments are gratefully acknowledged.
The financial support of the Ministry for Education, University and Research
(MIUR) is kindly acknowledged. 
%________________________________________________________________________

%______________________________________________________________________

%______________________________________________________________________

%______________________________________________________________________

\newpage
\begin{figure}
%\centerline{\psfig{figure=fig1.ps,width=16cm}}
\plotone{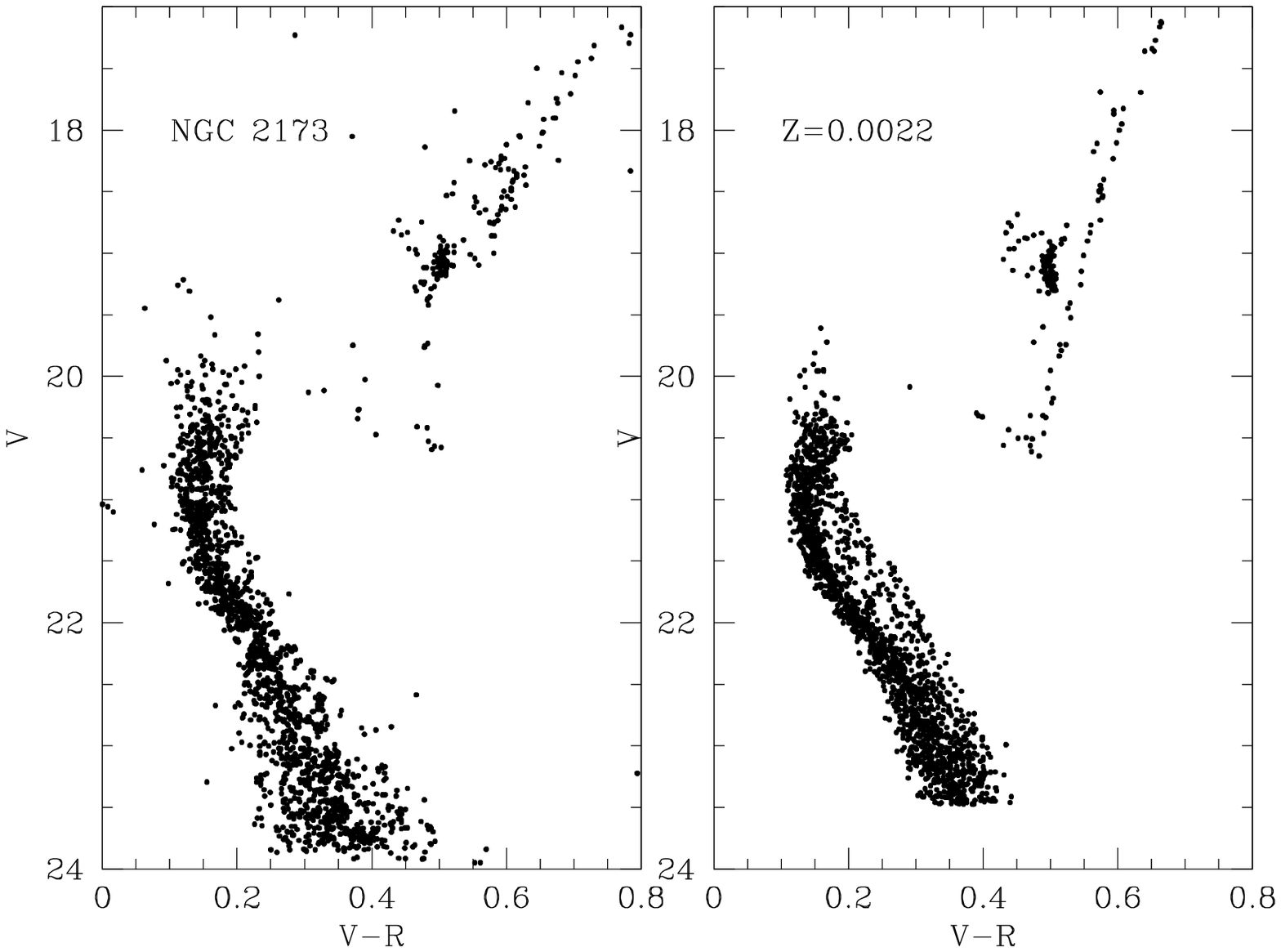}
\caption{NGC~2173. Left panel: the cluster color-magnitude diagram from VLT
data. Right panel: synthetic diagram for Z=0.0022 with a constant SFR  in
the age interval 1.4-1.7 Gyr 
}
\label{fig1}
\end{figure}

\newpage
\begin{figure}
%\centerline{\psfig{figure=fig2.ps,width=6cm}
\plotone{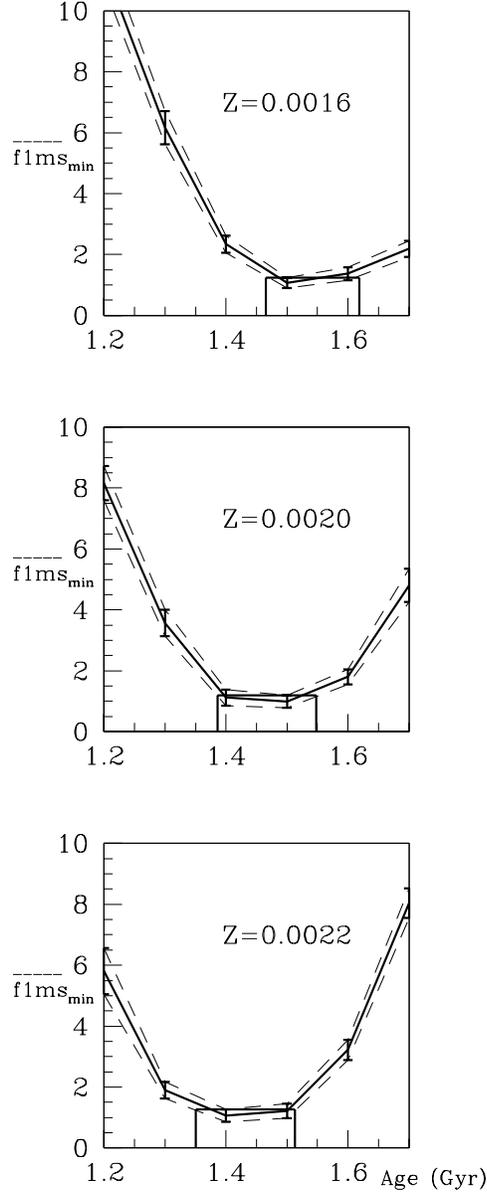}
\caption{The test.
Trend of the
average $f1ms_{min}$ (solid line) as a function of the final age of the 
star formation process for the metallicities Z=0.0016, 0.0020 and 0.0022.
The dashed lines represent the average $f1ms_{min} \pm 1 \sigma$.
 The thicker  vertical
bars identify the interval of age corresponding to the evaluated uncertainty.
 }
\label{fig2}
\end{figure}

\newpage
\begin{figure}
%\centerline{\psfig{figure=fig3.ps,width=8cm}
\plotone{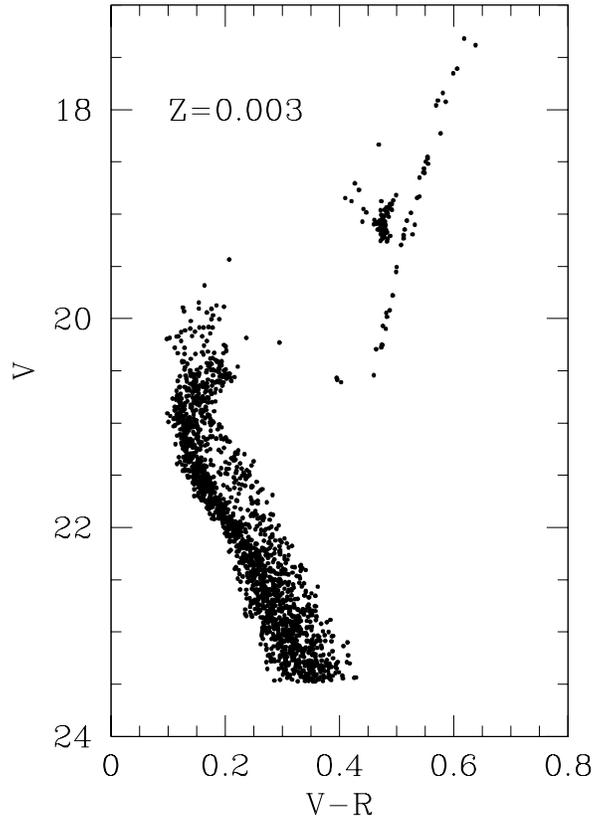}
\caption{Synthetic diagram of NGC~2173 for Z=0.0030 and a constant SFR  in
the age interval 1.5-1.8 Gyr (a solution of the partial criterium)}
\label{fig3}
\end{figure}

\newpage
\begin{figure}
%\centerline{\psfig{figure=fig4a.ps,width=16cm}}
\plotone{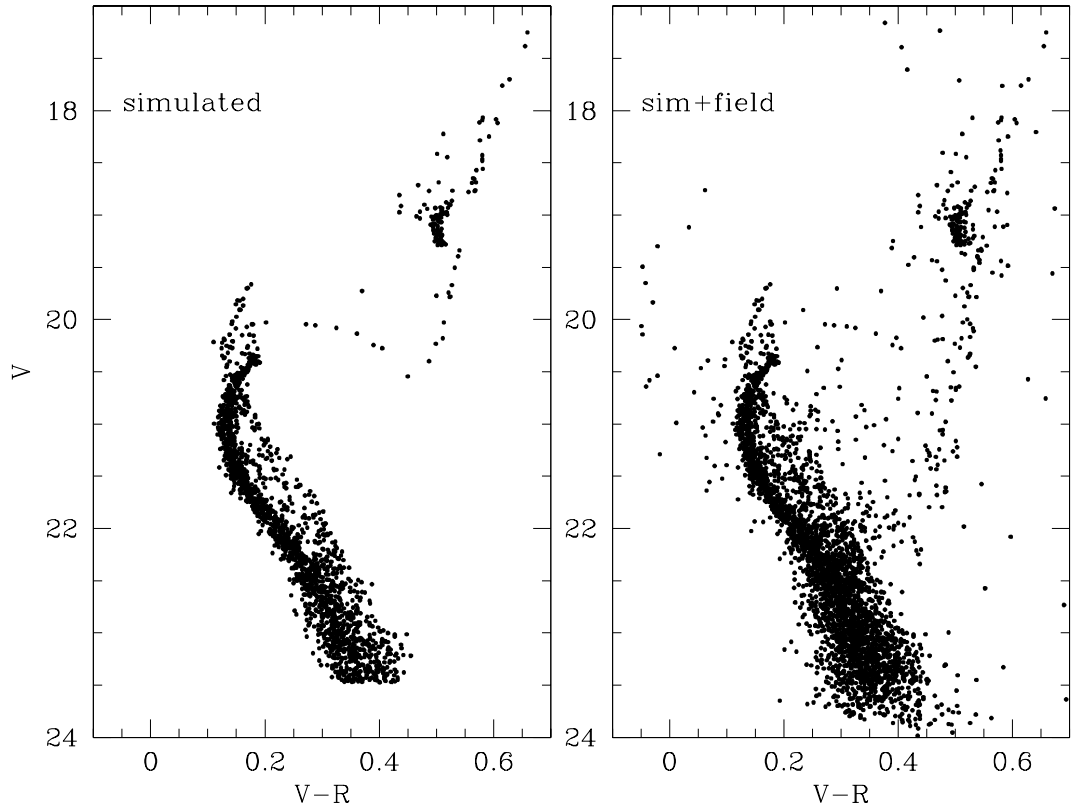}
\caption{a)
Left panel: synthetic diagram for the age 1.5 Gyr, 
metallicity Z=0.0022 and $30 \% $ binaries.
 Right panel: synthetic CMD  to which the field 
has been added.  
}
\label{fig4a}
\end{figure}

\addtocounter{figure} {-1}
\newpage
\begin{figure}
%\centerline{\psfig{figure=fig4b.ps,width=16cm}}
\plotone{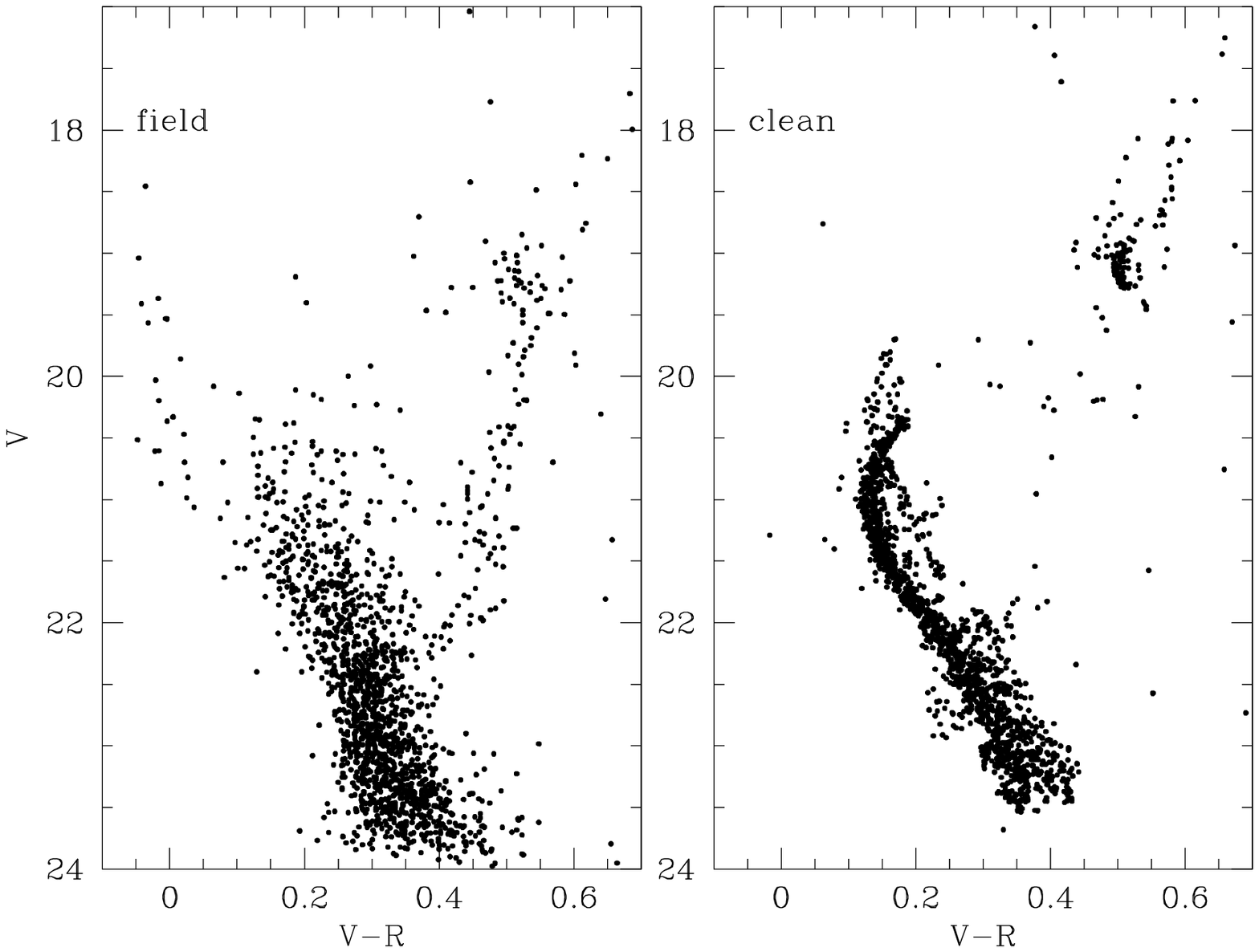}
\caption{b) 
Left panel: "control field" population used to 
decontaminate the "simulated cluster + field"  of the right panel in 
figure 4a.  Right panel: clean simulated cluster after the removal of the 
background contamination.   
}
\label{fig4b}
\end{figure}

\newpage
\begin{figure*}
%\centerline{\psfig{figure=fig5.ps,width=16cm}}
\plotone{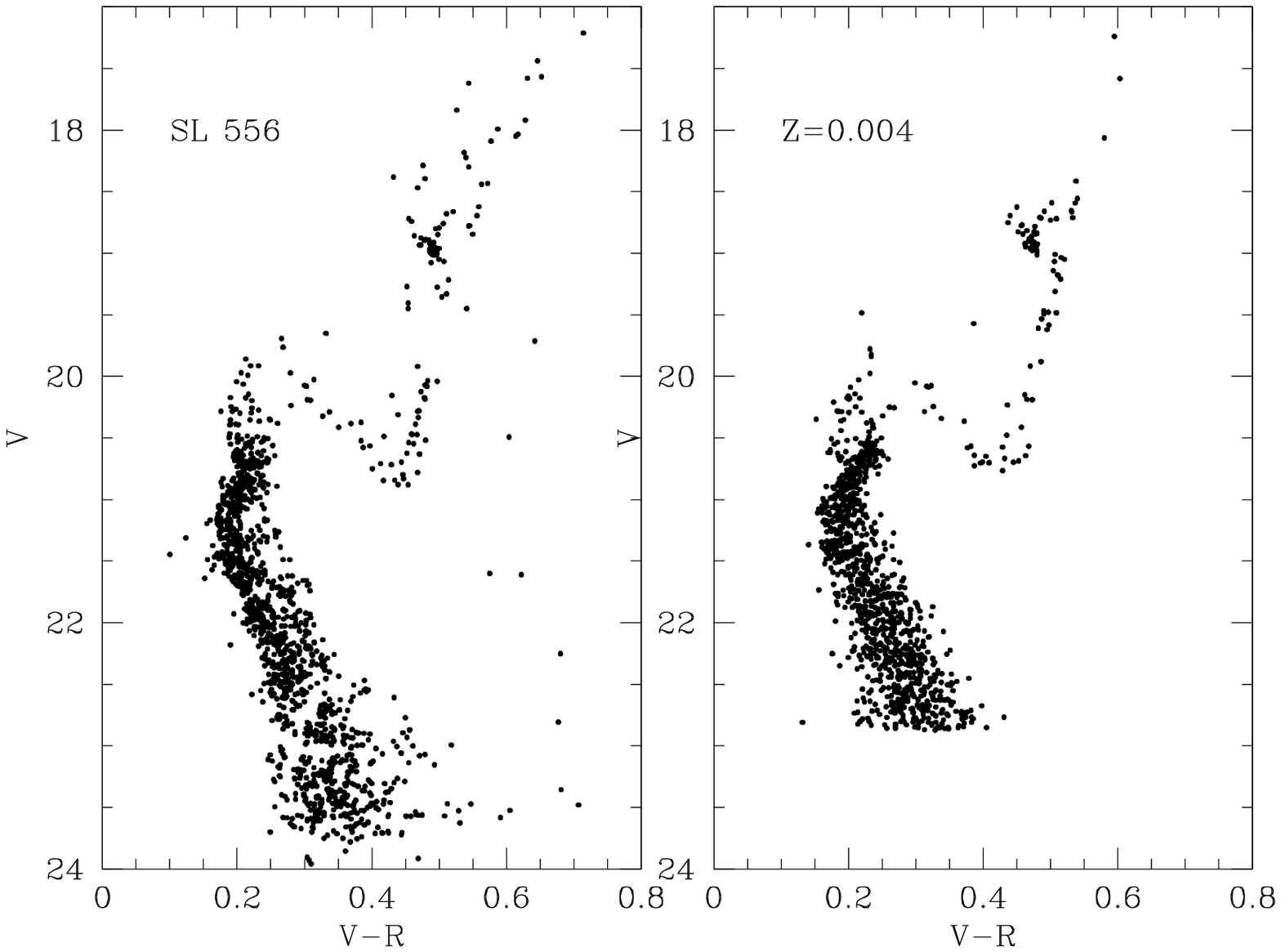}
\caption{SL~556. Left panel: the cluster color-magnitude diagram from VLT
data. Right panel: synthetic diagram for Z=0.004 and age 2.0 Gyr 
}
\label{fig5}
\end{figure*}

\newpage
\begin{figure*}
%\centerline{\psfig{figure=fig6.ps,width=16cm}}
\plotone{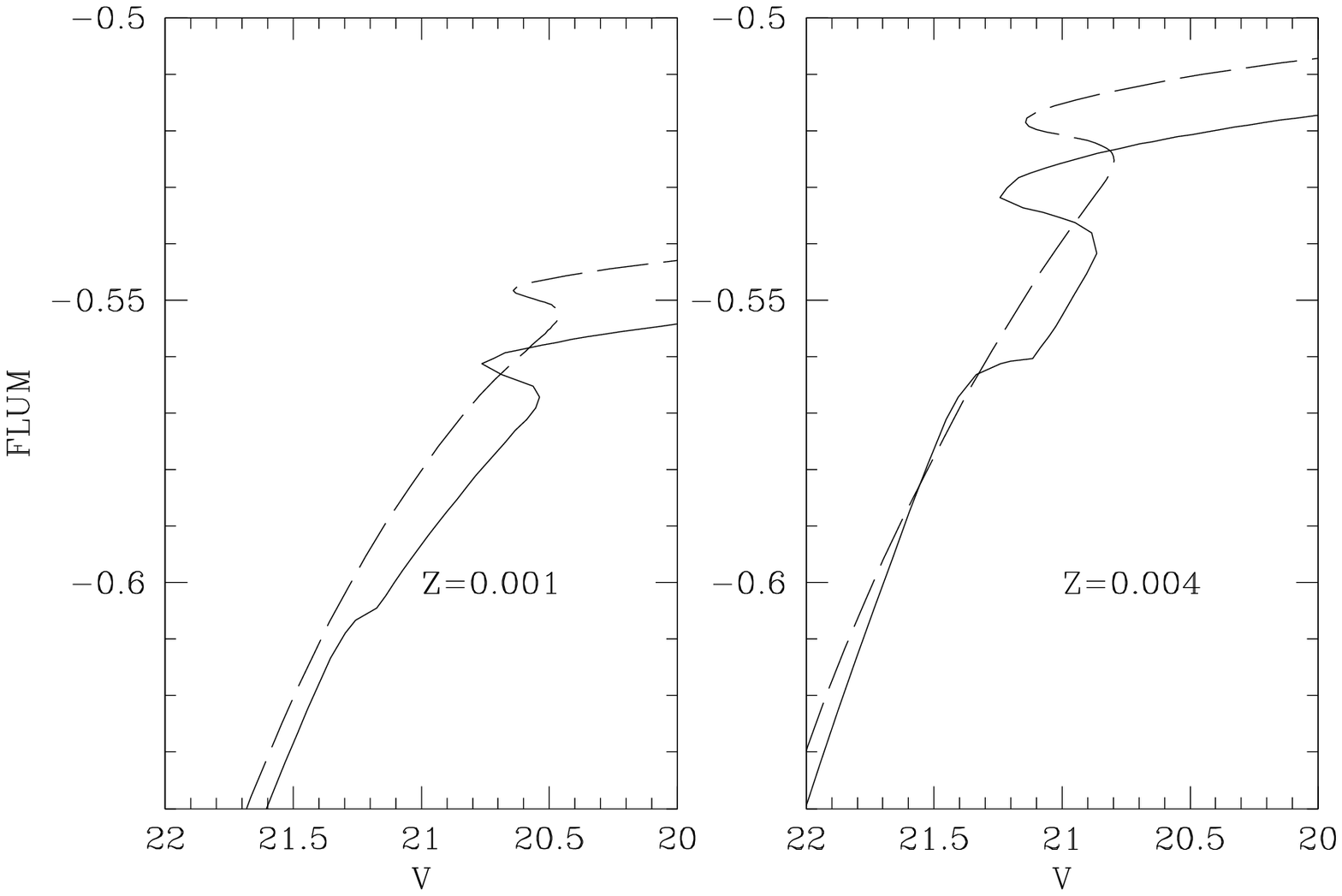}
\caption{Overshoot effect. Left panel: FLUM function (proportional to the 
number of stars in the corresponding mass interval) at the LMC distance,
age=3 Gyr and Z=0.001 (long-dashed line from $Y^2$ isochrones without
overshoot, solid line from Padova isochrones with overshoot). Right panel:
the same but for Z=0.004).
}
\label{fig6}
\end{figure*}

\newpage
\begin{figure*}
%\centerline{\psfig{figure=fig7.ps,width=16cm}}
\plotone{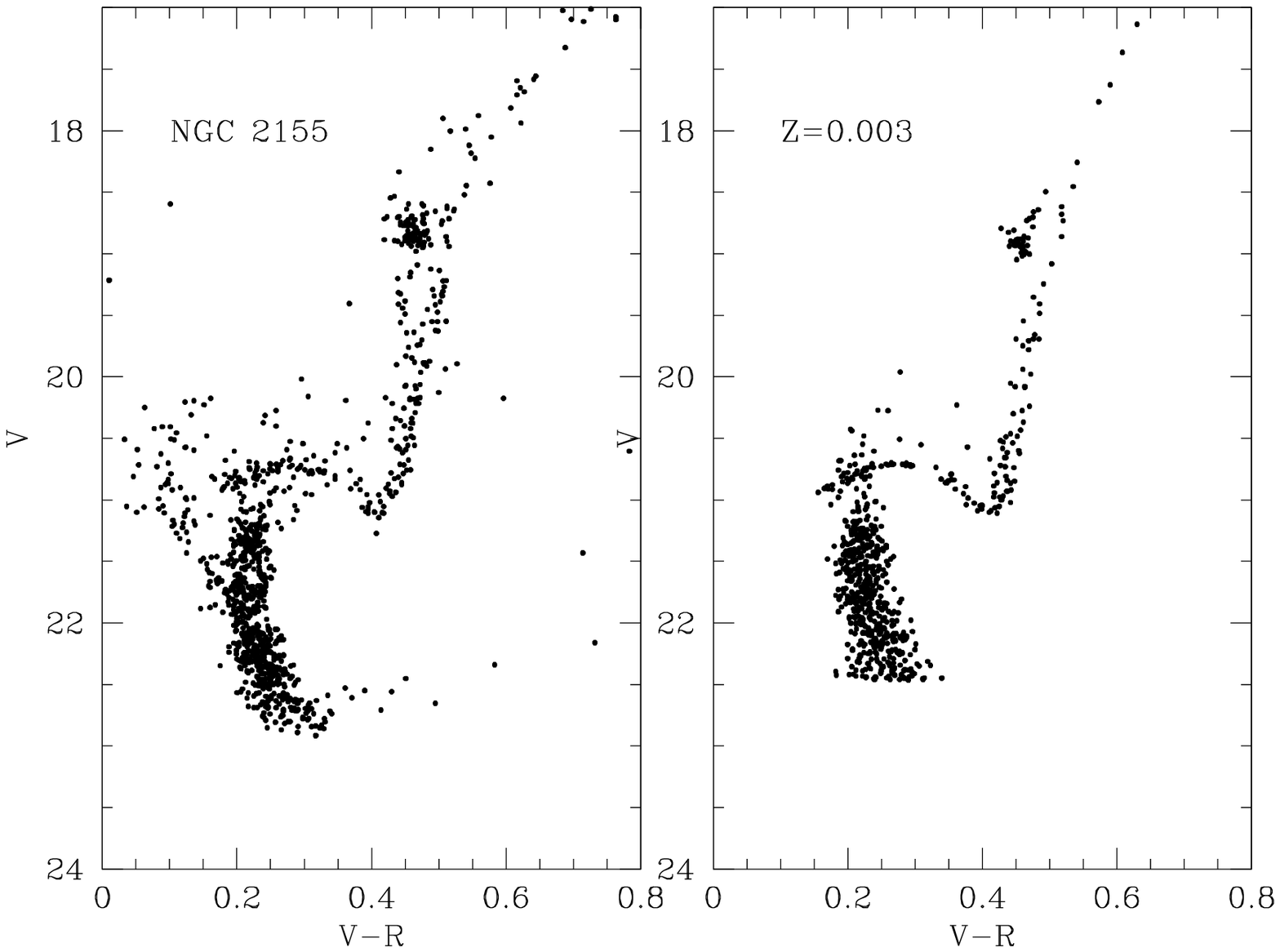}
\caption{NGC~2155. Left panel: the cluster color-magnitude diagram from VLT
data. Right panel: synthetic diagram for Z=0.003 and age 2.8 Gyr.
}
\label{fig7}
\end{figure*}

%______________________________________________________________________

%______________________________________________________________________

%______________________________________________________________________

\end{document}